\documentclass[twoside,leqno,twocolumn]{article}

\usepackage[letterpaper]{geometry}

\usepackage{ltexpprt}

\usepackage{graphicx,epstopdf}
\usepackage{subfigure} 
\usepackage{booktabs}  
\usepackage{multirow} 
\usepackage{multicol} 

\usepackage{mathtools}   
\usepackage{amsfonts}  

\usepackage{enumerate}

\usepackage{paralist}

\begin{document}

	\title{Cognitive Diagnosis with Explicit Student Vector Estimation and Unsupervised Question Matrix Learning}
	\author{Lu Dong\footnotemark[1]
		\and Zhenhua Ling\footnotemark[1]
		\and Qiang Ling\footnotemark[1]
		\and Zefeng Lai\footnotemark[1]}
	
	\date{}
	
	\maketitle
	
		\renewcommand{\thefootnote}{\fnsymbol{footnote}}
\footnotetext[1]{University of Science and Technology of China, \{dl1111, lzf1402\}@mail.ustc.edu.cn, \{zhling, qling\}@ustc.edu.cn}

	
	
	
	
	

	\begin{abstract} \small\baselineskip=9pt 
		Cognitive diagnosis is an essential task in many educational applications.
		Many solutions have been designed in the literature.
		The \textit{deterministic input, noisy ``and" gate} (DINA) model is a classical cognitive diagnosis model and can provide interpretable cognitive parameters, e.g., student vectors.
		However, the  assumption of the probabilistic part of DINA is too strong, because it assumes that the  slip and guess rates of  questions are student-independent.
		Besides, the question matrix (i.e., $Q$-matrix) recording  the skill distribution of the questions  in the cognitive diagnosis domain    
		often requires precise labels given by domain experts.
		Thus, we propose an \textit{explicit student vector estimation} (ESVE) method to estimate the student vectors of DINA with a local self-consistent test, which does not rely on any  assumptions for the probabilistic part of DINA.  
		Then, based on the estimated student vectors, the probabilistic part of DINA can be modified to a student-dependent model that the slip and guess rates are related to student vectors.
		Furthermore, we propose an unsupervised  method called \textit{heuristic bidirectional calibration algorithm} (HBCA) to label the
		$Q$-matrix automatically, which connects the question difficulty relation and
		the answer results for initialization and uses the fault tolerance of ESVE-DINA for calibration. 
		The experimental results on two real-world datasets show that   ESVE-DINA outperforms the  DINA model on  accuracy and that the $Q$-matrix labeled automatically  by HBCA can achieve performance comparable to that obtained with the manually labeled $Q$-matrix when using the same model structure. 
	\end{abstract}
	
	
	\section{Introduction}
	%
	Recently, many studies have been devoted to  computer-aided applications, e.g., computer-adaptive tests \cite{gershon2005computer, nichols2012cognitively}, teaching plan improvements \cite{chen2017tracking} and personalized  recommendation \cite{templin2006measurement}.
	Among these applications, cognitive diagnosis used to diagnose a student's degree of mastery of knowledge \cite{decarlo2011analysis, wu2015cognitive} is an essential task. 
	Moreover, it is important to note that the effectiveness of cognitive analysis is usually validated by the 
	\textit{predicting examinee performance} (PEP) task, which utilizes trained cognitive parameters from the previously	obtained examinee responses to predict unseen scores.
	
	Many cognitive diagnosis models (CDMs) have been developed to define cognitive parameters and progress, such as MIRT \cite{chalmers2012mirt}, DINA \cite{junker2001cognitive,decarlo2011analysis} and FuzzyCDF \cite{wu2015cognitive}. 
	CDMs assume that examinees can be characterized by the proficiency on specific skills, where a \textit{$Q$-matrix} \cite{rupp2008effects} denotes the skill distribution of all questions, and the skill distribution of a question composes one binary \textit{question vector}. The $Q$-matrix is a key feature of the question database. However, the annotation of
	the $Q$-matrix is always arduous \cite{gu2018sufficient,wang2020q}, because experts need to extract  specific knowledge of each question, which requires professional  abstractions.
	For another, the \textit{deterministic input, noisy ``and" gate model} (DINA) model is a well-known  CDM baseline due to its interpretive student vector parameters \cite{wu2015cognitive, wang2019neural}, and it is composed of a cognitive part and a probabilistic part. The cognitive part assumes that a student can answer a question correctly in theory if he or she masters all the skills tested by the question. Here, the student's mastery degrees of all  skills comprise a binary \textit{student vector}. 
	Moreover, its probabilistic part represents the students' scores with  slip and guess rates. Both of these two parts of DINA are  explanatory. However, DINA  utilizes a  strong assumption that the slip and guess rates of each question are student-independent, 
	%
	which reduces the  complexity of modeling but  goes against common sense. For example, elementary students and college students can have different guess rates on  the same college questions. 
	
	In this paper, we propose an \textit{explicit student vector estimation} (ESVE) method to estimate  student vectors of DINA locally  without any assumptions for its probabilistic part. Specifically, we filter slipped or guessed questions by testing the self-consistency of the question vectors with  answer labels,
	which only requires the cognitive part of DINA.   Next, ESVE-DINA estimates student vectors with their bounds from the  remaining questions that are not guessed and slipped.
	Then, based on the estimated student vectors, the probabilistic part of DINA can be modified to a student-dependent model that the slip and guess rates are related to student vectors.
	%
	Furthermore, we propose a \textit{heuristic bidirectional calibration algorithm} (HBCA) to label the $Q$-matrix automatically  with an initialization method and a bidirectional calibration process.
	First, the $Q$-matrix is initialized using a heuristic assumption that  the relatively easier questions examine fewer skills. 
	Then, we obtain these relative difficulty from the answer results and label questions  by taking relatively easier questions as bases.
	Additionally, we find that the self-consistency test of ESVE-DINA can circumvent  the errors of  $Q$-matrix. Then,  a \textit{dual algorithm} (DA) of ESVE-DINA is designed to  estimate the $Q$-matrix from the estimated student vectors, which is a dual task of student vector estimation from the $Q$-matrix. Thus, the $Q$-matrix can be bidirectionally calibrated by conducting the fault tolerance of both  ESVE-DINA and DA.  
	The main contributions of this paper are summarized as follows:
	\begin{itemize} 
		\item We propose an ESVE algorithm to estimate the student vectors of
		 DINA, which requires no  assumption for its probabilistic part.  Moreover,  the student-independent probabilistic part of DINA can be modified to a student-vector-related model based on the estimated student vectors.
		
		%
		\item  We also propose an unsupervised  method HBCA to label the $Q$-matrix automatically,  which connects  the question difficulty relation and the answer results for initialization and uses the fault tolerance of ESVE-DINA for calibration. 
		\item Experiments on two real-world datasets \textit{Fraction} and \textit{ASSISTments2015} show that ESVE-DINA outperforms DINA model on accuracy, and the $Q$-matrix labeled automatically by HBCA can achieve  performance comparable to that obtained  with manual $Q$-matrices  using the same model structure. %
	\end{itemize} 
	
	\section{RELATED WORK}
	We briefly summarize our related work for cognitive diagnosis from
	two aspects: cognitive diagnosis methods and the question information annotation domain.

	\subsection{Cognitive Diagnosis Methods}
	In educational psychology, 
	a fundamental  CDM is the \textit{deterministic inputs, noisy ``and" gate} (DINA)  model \cite{junker2001cognitive,haertel1984application,de2011generalized}. It assumes that a student can answer a question correctly when he or she masters all the tested skills. Next, he or she may slip or guess this question after the ideal process. 
	Though these  parts are both reasonable, 
 DINA utilizes a strong assumption that the slip and guess rates are student-independent, decoupling the student vectors and the slip and guess rates to achieve an acceptable complexity  of modeling. However,
	 ESVE method  can estimate student vectors with only the cognitive hypothesis of DINA, which avoids this restriction.
	\subsection{Question Information Annotation}
	The problem of question information annotation is often polarized and is generally either finely carried out by  experts in a costly manner or coarsely labeled with fuzzy annotations. For one,  as shown in \cite{gu2018sufficient,wang2019neural}, labeling of a detailed $Q$-matrix such as the dataset \textit{Fraction} \cite{tatsuoka1984analysis} requires  domain experts with abstractions and  is quite costly.
	For another, many question datasets have sparse skill labels, such as the family of \textit{ASSISTments} datasets\renewcommand{\thefootnote}{1}\footnote{https://sites.google.com/site/assistmentsdata/home}. Its  questions usually have 1 to 3 sparse and nonspecific skill labels, notably increasing  the  burden of modeling. In this paper, we show that
	the initialization of the $Q$-matrix can be related to the answer results, and the fault tolerance of the solving algorithm can calibrate the initialized $Q$-matrix.
	
	\section{Problem Definition}
	Here, we will introduce the formal definition of cognitive diagnosis  and the DINA model.
	
	We study cognitive diagnosis for  cognitive parameters $V$ of S students on M questions with question parameters $B$. Given the existing answer results $X$, skill distribution of questions $Q$-matrix and score model CDM $X = f(V, B, Q)$, we need to solve student parameters $V$  and  question parameters $B$, which can also be evaluated by the PEP task of predicting unseen $X$ to show the rationality of a CDM and  the estimated $V$. 
	
	
	Let us first review the DINA model \cite{junker2001cognitive,decarlo2011analysis,von2014dina} 
	\begin{eqnarray}
	&\xi_{ij}  = \prod_{k=1}^N \alpha_{ik}^{Q_{jk}}.\label{e3.1} \\
	&s_j  = P(X_{ij}=0|\xi_{ij}=1).\label{e3.2} \\
	&g_j  = P(X_{ij}=1|\xi_{ij}=0). \label{e1.3} \\
	&P(X_{ij}=1|\xi_{ij})  = (1-s_j)^{\xi_{ij}}g_j^{1-{\xi_{ij}}}. \label{e1.4}
	\end{eqnarray} 
	DINA  uses a binary latent variable $\xi_{ij}$ to denote the  cognitive part. Ideally, $i$-th student  can answer $j$-th question  correctly when he or she masters all the test  skills. If there exists a $k$-th skill  that is tested ($Q_{jk}=1$) but  is not mastered 
	 ($\alpha_{ik}=0$),  $\xi_{ij}$ is false. $\boldsymbol Q_{j}$ stands for the binary question vector of the $j$-th question, and $\boldsymbol \alpha_{i}$  is the binary  student vector of the  $i$-th student.  The probabilistic part of DINA uses  $s_j$ to denote the slip rate of $j$-th question, i.e., the probability that the ideal result of the $j$-th question is correct but is answered incorrectly. Then, the probability of $X_{ij}$ being true is $1 - s_j$. $g_j$ denotes the guess rate in a similar manner. However, every $s_j$ and $g_j$ are student-independent for their independence on the student index $i$.

	
	\section{DINA with Explicit Student Vector Estimation}
	
	In this section, we will introduce our ESVE method, which filters slipped and guessed questions by a local self-consistency test of the $Q$-matrix with answer labels. This indicates that we will derive each student vector $\boldsymbol  \alpha_{i}$ with only Eq. (3.1) but no probabilistic assumption for $s_j$ and $g_j$ of DINA, while traditional methods \cite{decarlo2011analysis} optimize the joint probability Eq. (3.4) of all students.
	
	
	ESVE is a two-step method, where in the first stage, the observed case is converted  to an ideal case in which there are no slipped or guessed questions, and in the second stage, the  feasible student vector is estimated from the ideal case. 
	Then, primarily,  we infer some ideal intermediate relations when there are no slipped or guessed questions; i.e., for all $j$, $s_j=g_j=0$.
	
	First, we define some intermediate variables. 
	\begin{Definition}
		If a student did a question set, then the question answered correctly by guessing or incorrectly by slipping is called an \textit{unreliable question}, and the corresponding question vector is an unreliable question vector. Otherwise, the question is called a \textit{reliable question} and  corresponds to a reliable question vector.
	\end{Definition}
	%
	Meanwhile,   the question vector set  are divided into two subsets by the answer results $X_i$ of each student $i$,
 the correct question set and the incorrect question set; their corresponding question vector sets are annotated as $Q^T$ and $Q^F$ respectively. 
	
	Then,  we rewrite   Eq. (3.1)   as 
	\begin{equation}\label{e4.5}
	X_{ij}= \xi_{ij} = \left\{
	\begin{array}{ll}
	1, & \mbox{if for all k, }\alpha_{ik} \ge Q_{jk}, \\
	0, & \mbox{otherwise}.
	\end{array}
	\right.
	\end{equation}
	Here, $X_{ij}=\xi_{ij}$ is inferred by Eq. (3.4) with the ideal case that for all $j$, $s_j=g_j=0$. The second equation means that the $i$-th student  can answer the $j$-th question  correctly once he or she masters all the  tested skills.  Eq. (4.5) has  the same output as Eq. (3.4) for identical input. 
	
	Thus, based on Eq. (4.5), ideally, $X_{ij}=0$ means that there exists at least one $k$, $\alpha_{ik} < Q_{jk}$. Similarly, $X_{ij}=1$ means that for all $z$, $\alpha_{iz} \ge Q_{jz}$ . When considering all the question vectors that are divided into $Q^T$ ($X_{ij}=1$) and $Q^F$ ($X_{ij}=0$), we obtain that
	\begin{eqnarray}
	&\forall Q_p^T \in Q^T, \forall z,  
	\alpha_{iz} \ge Q_{pz}^T. \\
	&\forall Q_q^F \in Q^F, \exists k, \mbox{ s.t. } \alpha_{ik} < Q_{qk}^F.
	\end{eqnarray}
	
	In the following two subsections, we first infer the conflict degrees of the question vectors with observed answer labels to filter unreliable questions,  converting the observed answer results to only ideal reliable questions. Second, we induce a feasible student vector satisfying all bounds from the remaining reliable questions.
	
	
	
	
	%
	\subsection{Filtering Unreliable Question Vectors}
	In this subsection, we describe  a method called \textit{conflict detection} to obtain  conflict degrees, which  represent the  unreliable degrees of \textbf{question vectors with observed result labels} ($Q^T, Q^F$). Then, we filter all unreliable questions based on conflict degrees. 
	
	Here, we will analyze the relationship between the  question vector pairs  with  observed right and wrong labels ($Q_p^T$ and $Q_q^F$). 
	
	First,  a globally reliable condition between $Q^T$ and $Q^T$  can be inferred by  the ideal conditions in Eq. (4.6) and Eq. (4.7). From these two equations, by setting  $z$ as a specific $k$, we obtain that 
	\begin{eqnarray} 
	& \hspace{+4mm} \forall Q_p^T, \forall Q_q^F, 
	\exists k, \mbox{s.t. }  Q_{pk}^T \leq \alpha_{ik} < Q_{qk}^F.
	\end{eqnarray}
	Then, we can  find a locally self-consistent condition between each $Q_p^T$  and $Q_q^F$ pair as
	\begin{eqnarray}
	\exists k, \mbox{ s.t. }  Q_{pk}^T < Q_{qk}^F.
	\end{eqnarray}
	Here, the local self-consistent condition between  $Q_p^T$ and  $Q_q^F$ means that both of them are more reliable locally, because these two question
	vectors satisfy a part of the whole reliable conditions in Eq. (4.8).   Thus, an approach to identify the  unreliable questions is to count the question vectors that break the globally self-consistent condition in Eq. (4.8) the most. The reverse condition of Eq. (4.9)  becomes
	\begin{eqnarray}
	\forall k,Q_{pk}^T \ge Q_{qk}^F .
	\end{eqnarray}
	The condition in Eq. (4.10) for detecting each right and wrong question  pair is called  the \textit{conflict condition}. Intuitively, we should treat right and wrong questions equally, which means that if a $Q_p^T$ and $Q_q^F$ pair satisfy Eq. (4.10), the conflict degrees of question $q$ and $p$ should be  increased by one  together.  Then, we can consider the conflict degrees to be the representation of  unreliable degrees, and the questions with the maximum conflict degrees will be filtered because they  break the globally reliable condition in Eq. (4.8) the most.
	
	Thus, we can count the conflict degrees of all questions by traversal detection of every right and wrong question pairs using Eq. (4.10). We call this \textit{conflict detection}. 
	Moreover, we can convert  observed questions to reliable questions by conflict detection and filtering questions with maximum conflict number until there is no conflict.  Here, we can observe that the filtering progress uses no probabilistic assumption outside of the cognitive part of DINA, and then, it avoids the strong probabilistic assumption of DINA.
	
	\subsection{Estimating Student Vector from Reliable Question Vectors}
	In this subsection, we estimate every student vector from \textbf{reliable question vectors} ($Q_{rel}^T, Q_{rel}^F$) by estimating the upper and lower bounds of its component, which means that this is an ideal case shown in \S 4. We still use $Q^T, Q^F$  for simplicity.  
	
	From Eq. (4.6), we obtain that all  $Q_{pz}^T$ is not larger than $\alpha_{iz}$, and since the relation in Eq. (4.6) is for all components $z$, we can set $z$ equal to $k$ for the unity of derivation, and we call the union of all components $Q_{pk}^T$  the \textit{lower component} of $\alpha_{ik}$. Thus, we obtain that
	\begin{eqnarray}
 \alpha_{ik } \ge \alpha_{ik}^{lower} = \bigcup_{p=1}^{u} Q_{pk}^T.
	\end{eqnarray}
	Here, the OR operation means that student $i$ should master skill $k$ if the right question set tests skill $k$, and $\alpha_{ik}^{lower}$  can be thought as the maximum value of all $Q_{pk}^T$.  $u$ denotes the size of the right question set.
	
	As there are no unreliable questions, the  whole $Q$-matrix and student vector $\boldsymbol \alpha_{i}$ must be self-consistent. Thus, from Eq. (4.11) and Eq. (4.7), we obtain that
	\begin{eqnarray}
	\hspace{+8mm} \forall  Q_q^F \in Q^F, \exists  k,\mbox{ s.t. } Q_{qk}^F > \alpha_{ik} \ge \alpha_{ik}^{lower}. 
	\end{eqnarray}
	As  $\alpha_{ik}$ is unknown, Eq. (4.12) can be written as 
	\begin{eqnarray}
	%
	\forall Q_q^F \in Q^F, \exists k, \mbox{ s.t. } Q_{qk}^F > \alpha_{ik}^{ lower}.
	\end{eqnarray}
	Likewise, we can consider the \textit{upper component} of $\alpha_{ik}$, which is defined as a component that is larger than $\alpha_{ik}$, which comes from $Q_{qk}^F$ in Eq. (4.12). As  $\alpha_{ik}$ is unknown, we  use  Eq. (4.13) to infer the upper component of $\alpha_{ik}$,  which implies that
	\begin{eqnarray}
	\alpha_{ik}^{upper} = \bigcap_{q=1}^{v} Q_{qk}^F, \mbox{if } Q_{qk}^F > \alpha_{ik}^{lower}.
	\end{eqnarray}
	
	Here, the AND operation means taking the minimum values by the definition of the upper components. However, this operation is not unique. For example, consider the case of only two reliable questions, a wrong question vector  $[1,1,1]$ and a right question vector $[0,0,1]$. We can see that the first and the second components can both satisfy Eq. (4.14), so they are all upper components. But there may be three  reasons why the student answered the question q incorrectly; namely, he or she  lacks skill 1 or  skill 2 or  both. 
	Nevertheless, to  punish wrong questions, we  choose the third-worst solution, i.e., choosing the worst combination of the observed lacked skill set.
	\begin{figure}[htbp]
		\centering
		\includegraphics[scale=0.4]{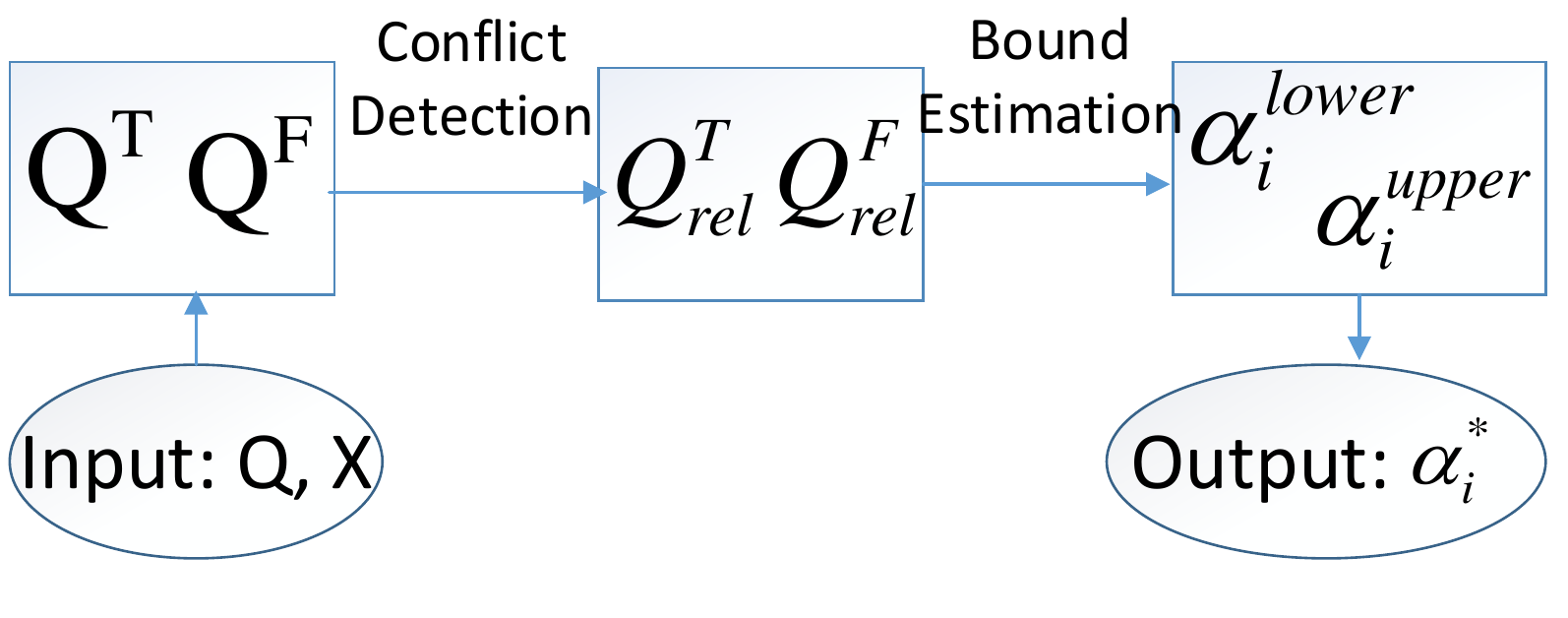}
		\caption{Flowchart of ESVE-DINA.}
	\end{figure}

	Next, a feasible $\alpha_{ik}$ satisfying the upper and lower components of $\alpha_{ik}$ will be selected.
	%
	%
	%
	%
	The upper and lower bounds  of each $\alpha_{ik}$ are inferred by Eq. (4.11) and Eq. (4.14) from reliable question vectors. Then, we just choose a feasible $\alpha_{ik}$ that meets the condition $\alpha_{ik }^{lower} \leq \alpha_{ik} < \alpha_{ik}^{ upper} $. As $\alpha_{ik}$ is either 0 or 1,  $\alpha_{ik}$ is set as follows:
	\begin{equation}
	\alpha_{ik} = \left\{
	\begin{array}{ll}
	0, & \mbox{if }  \alpha_{ik }^{upper} =1;\\
	1,&  \mbox{if }\alpha_{ik }^{lower}=1; \\
	random(0,1), & \mbox{others}.
	\end{array}
	\right.
	\end{equation}
	
	Therefore,  every student vector $\boldsymbol \alpha_{i}$ are obtained by the same operation on each $\alpha_{ik}$.
	
	\subsection{DINA with Explicit Student Vector Estimation}
	Combining \S 4.1 and \S 4.2, the student vectors of DINA are solved by ESVE. First, the $Q$-matrix are divided into $Q^T, Q^F$ by the observed answer results. Then, the unreliable questions are filtered by conflict detection using Eq. (4.10). 
	Next, the upper and lower bounds of each $\alpha_{ik}$ are estimated by the remaining reliable question vectors   using Eq. (4.11) and Eq. (4.14). Finally, every feasible  $\alpha_{ik}$ are obtained by Eq. (4.15).
	The flowchart of ESVE-DINA is shown in Fig. 1.
	
	
	
	Here, we can observe that  ESVE-DINA estimates student vectors with no probabilistic assumption outside of the cognitive part of DINA. Then,  the prediction method based on  ESVE-DINA can be quite flexible. 
	
	A trivial method is the student-independent (SI) prediction method of DINA, where $s_j, g_j$ are unrelated to the student vector $\boldsymbol \alpha_{i}$ or other possible parameters. $s_j, g_j$  are obtained by  their definitions. For example, $s_j$= N(question $j$ filtered from wrong question set)/N(examinee of question $j$). Here,  $N(X)$ means the number of $X$, and a question filtered from wrong question set means that it should be  right but in fact is  wrong, meaning it is  a slipped question. $g_j$ counts the number filtered from the right question set similarly. 
	
	A more reasonable method is that $s_j, g_j$ are student-dependent (SD) and they can be related to the student  vector $\boldsymbol \alpha_{i}$. We assume that $s_j $ is related to the mastery  skill
	number (i.e., level) of   $\boldsymbol \alpha_{i}$,  meaning that we think that students with the  same  skill number have identical slip rates on each question. Furthermore,  we assume that $g_j $ is related to the lacked skill number (i.e., deficiency) of some student  on  question $j$. SD $s, g$ are computed in a similar manner as above SI $s, g$, both the numerator and  denominator plus a condition that the level or deficiency is equal to a specific number. 
	
	\section{Heuristic Bidirectional Calibration Algorithm}
	In this section, we  introduce our HBCA method. HBCA is a bidirectional calibration (BC) process with an  initialization procedure. The initialization of the $Q$-matrix spans a question tree, and it heuristically  assumes that relatively easier questions  tests  fewer skills, which is called the \textit{question spanning tree} (QST)  algorithm.  
	Besides, the BC process repeatedly utilizes ESVE-DINA and its dual algorithm (DA). 
	The fault tolerance of ESVE-DINA and DA are used for calibration of the $Q$-matrix.
	The following subsections will introduce QST, DA and HBCA. 
	
	\subsection{Question Spanning Tree Algorithm}
	In this subsection, we show how to initialize the $Q$-matrix automatically. 
	We assume that the relatively easier questions examine fewer skills, and an example of QST with three questions is shown in Fig. 2. There are two steps. The first is to find the relations from the student answer results, and the second is to initialize a $Q$-matrix by spanning a question tree with these relations. 
	
	
	First, we discuss the relations between questions. Supposing that there are two questions, we can know that a \textbf{covering relation} such as $Q_3$ and $Q_1$ in Fig. 2 can be useful, which means that every component of one question vector  is equal to or greater than   another. 
	
	The covering relation can efficiently restrict the solution space of the $Q$-matrix. If question $w$ covers a question set with question vector set [$Q_1$, $Q_2$, ..., $Q_z$], then we can obtain the following  inequality:
	\begin{align}
	\forall k, Q_{wk} \ge \bigcup_{p=1}^z Q_{pk}.
	\end{align}
	Here, the OR operation takes the maximum value.
	\subsubsection{Covering Relation Construction}
	In this subsection, we demonstrate a method for obtaining covering relations.
	First, the conditional probability of the question's accuracy  is defined and can be computed by counting statistics.
	\begin{align}
	\beta_{wz} = P(X_{iz}=1 | X_iw=1).
	\end{align}
	Based on our assumption that relatively easier questions test fewer skills, if $\beta_{wz}$ is very large; i.e., $\beta_{wz} \ge \eta$ and $\eta$ is large, then question $z$ is relatively easier than question $w$ and we think question $w$ covers question $z$. 
	
	\begin{figure}[htbp]
		\centering
		\includegraphics[scale=0.4]{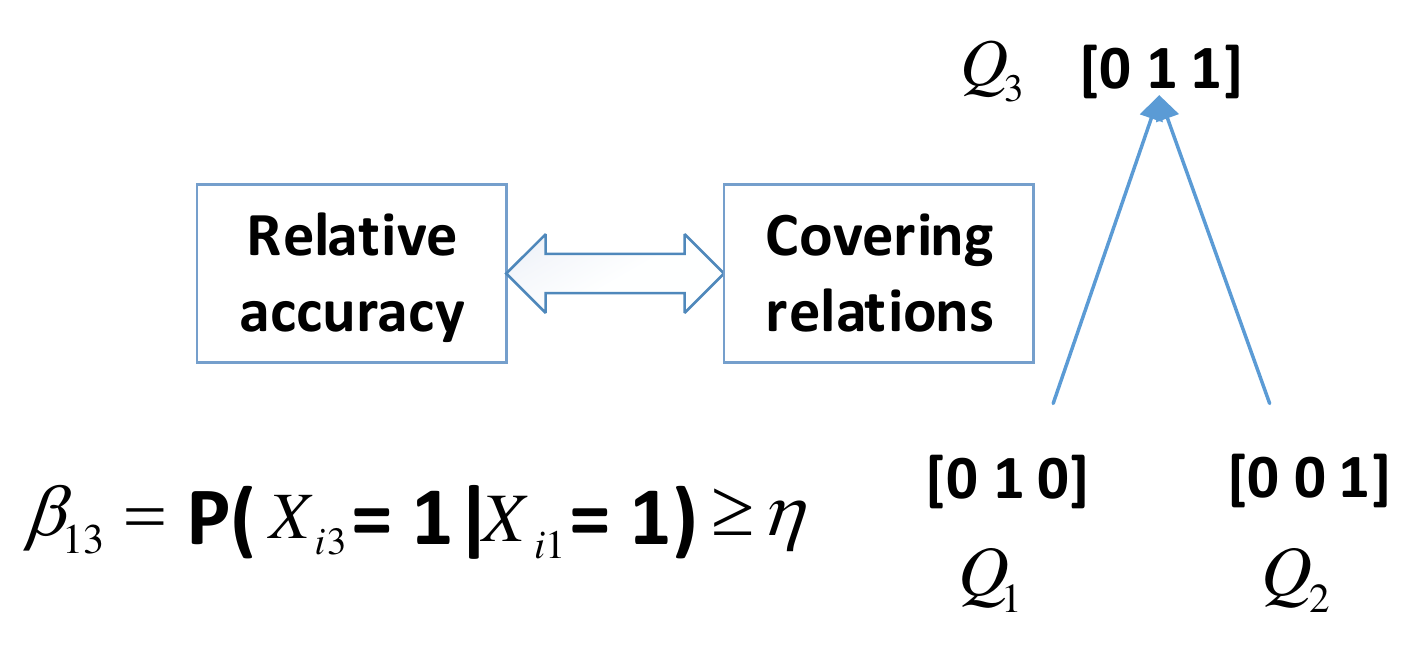}
		\caption{An example of QST.}
	\end{figure}
	%
	Moreover, from the standpoint of entropy \cite{gray2011entropy}, a large $\eta$  means that the answer of question $z$ severely tracks question $w$ and it is a compact piece of information. Meanwhile, the  $\eta$ value  can be guided by the generated $Q$-matrix.  Once it is too small, the whole covering relation will so redundant that the generated $Q$-matrix will have many full binary question vectors.
	
	\subsubsection{$Q$-matrix Initialization}
	In this subsection, we describe the initialization of the $Q$-matrix with covering relations.
	First, the question pair satisfying covering relations  are called a  parent and child pair. 
	As parent and child are the terms of tree structure \cite{pettie2002optimal}, we consider that our method spans the question tree and we call it the \textit{question spanning tree} algorithm. 
	
	Next, we  show how to initialize the $Q$-matrix with the  parent and child relation. From Eq. (5.16), we  know that if the children  of question $w$ is question $1$ to question $z$,  $Q_{wk}$  can be set  as its lower bound $\bigcup_{p=1}^z Q_{pk}$, such as $Q_3$ and $Q_1,Q_2$ in Fig. 2. Meanwhile, to make some randomness of generated $Q_w$, we can set some probability to roll over its zero component. 
	%
	For example, $Q_3$ in Fig. 2 may become [1, 1, 1] instead of [0, 1, 1]. Furthermore, we can randomly initialize the leaf nodes without the children. Moreover, we can span the question tree in the descending order of the parent number, which means that we label  the questions from easy to difficult. Here, more parents means an easier question. Thus, the $Q$-matrix is initialized with our relative relation assumption and base questions, i.e., children questions. 

	\subsection{Dual Algorithm}
	In this subsection, we  introduce the fault tolerance of ESVE-DINA and estimate the $Q$-matrix from student information with DA.
	
	First,  we analyze the process of ESVE-DINA. ESVE-DINA filters unreliable questions by the self-consistency test of the question vectors with  answer labels. If a question vector has more wrong labels with fixed result labels (result labels are unrelated to $Q$), it may
 fail the self-consistency test  easily and  
 will not influence the estimation of student vectors. Hence, ESVE-DINA may eliminate some mistakes in the $Q$-matrix.
 Thus, if we can avoid some mistakes of estimated student vectors, we can  calibrate the $Q$-matrix by these  bidirectional fault tolerances. Fortunately, we can  design a dual algorithm of ESVE-DINA to achieve this goal by the duality between the student vector and the question vector. The duality is as follows:
	
	(1) The answers from  student $i$ depend on all question vectors and one unknown student vector $\boldsymbol \alpha_{i}$.
	
	
	If student $i$ can answer question $j$ correctly (T label), then, for all $k$, $\alpha_{ik} \ge$  $Q_{jk}$;	otherwise (F label), there exists at least one $k$ such that $\alpha_{ik} < Q_{jk}$.
	
	\textit{Ideal condition}: $\forall p \mbox{ and } q, \exists k, Q_{qk}^F > \alpha_{ik} \ge Q_{pk}^T$.
	
	(2) The answers to  question $j$ depend on all student vectors  and one  unknown question vector $\boldsymbol Q_j$.
	
	
	If question j can be answered by student $i$ correctly (T label), then, for all $k$, $Q_{jk} \leq$ $\alpha_{ik}$; otherwise (F label), there exists at least one $k$ such that $Q_{jk} > $ $\alpha_{ik}$.
	
	\textit{Ideal condition}: $\forall p \mbox{ and } q, \exists k, \alpha_{qk}^F < Q_{jk} \leq \alpha_{pk}^T$.
	
	From the above comparison, we observe that the ideal condition of the student vector and the question vector only  differ in terms of the inequality direction. Thus, 
	we can design a dual algorithm (DA) of ESVE-DINA to estimate the $Q$-matrix by the following correspondence.
	Suppose that $\alpha^T$ has $r$ elements and that $\alpha^F$ has $l$ elements.
	
\begin{enumerate}[(1)]
		\item 
	
	conflict condition of DA: $\forall k, \alpha_{wk}^T \leq  \alpha_{zk}^F.$ $\iff$ conflict condition of ESVE-DINA: $\forall k, Q_{pk}^T \ge  Q_{qk}^F.$
	\item  upper component of $Q_{jk}$: $ Q_{jk}^{upper}  = \bigcap_{w=1}^r \alpha_{wk}^T$ $\iff$  lower component $\alpha_{ik}$: $\alpha_{ik}^{lower} = \bigcup_{p=1}^u Q_{pk}^T$;
	\item  lower component of $Q_{jk}$: $Q_{jk}^{lower} =  \bigcup_{z=1}^l{\alpha}_{zk}^F$, $ \mbox{if } \alpha_{zk}^F < Q_{jk}^{upper}$. $\iff$ upper component of $\alpha_{ik}$: $\alpha_{ik}^{upper} = \bigcap_{p=1}^u Q_{qk}^F, \mbox{ if }  Q_{qk}^F > \alpha_{ik}^{lower}$;

		\end{enumerate}
	Every $Q_{jk}$ can be set by an  equation similar to Eq. (4.15) that satisfies $Q_{jk }^{lower}<Q_{jk} \leq Q_{jk}^{ upper}$.
	
	

	\subsection{Heuristic Bidirectional Calibration Algorithm}
	Based on \S 5.1 and \S 5.2, HBCA can label the $Q$-matrix automatically. First, the $Q$-matrix are performed by the heuristic initialization of  QST, and then,  the student vectors (S) and the $Q$-matrix are bidirectionally calibrated by  ESVE-DINA and DA. Furthermore, an optimization strategy is  initializing many $Q$-matrices and update them separately (total T iterations), and if the training does not decrease the prediction error, some new $Q$-matrices are  initialized  to replace those with bad property. This method imitates the update of  genetic algorithm  \cite{whitley1994genetic}. 
	\begin{figure}[htbp]
		\centering
		\includegraphics[scale=0.4]{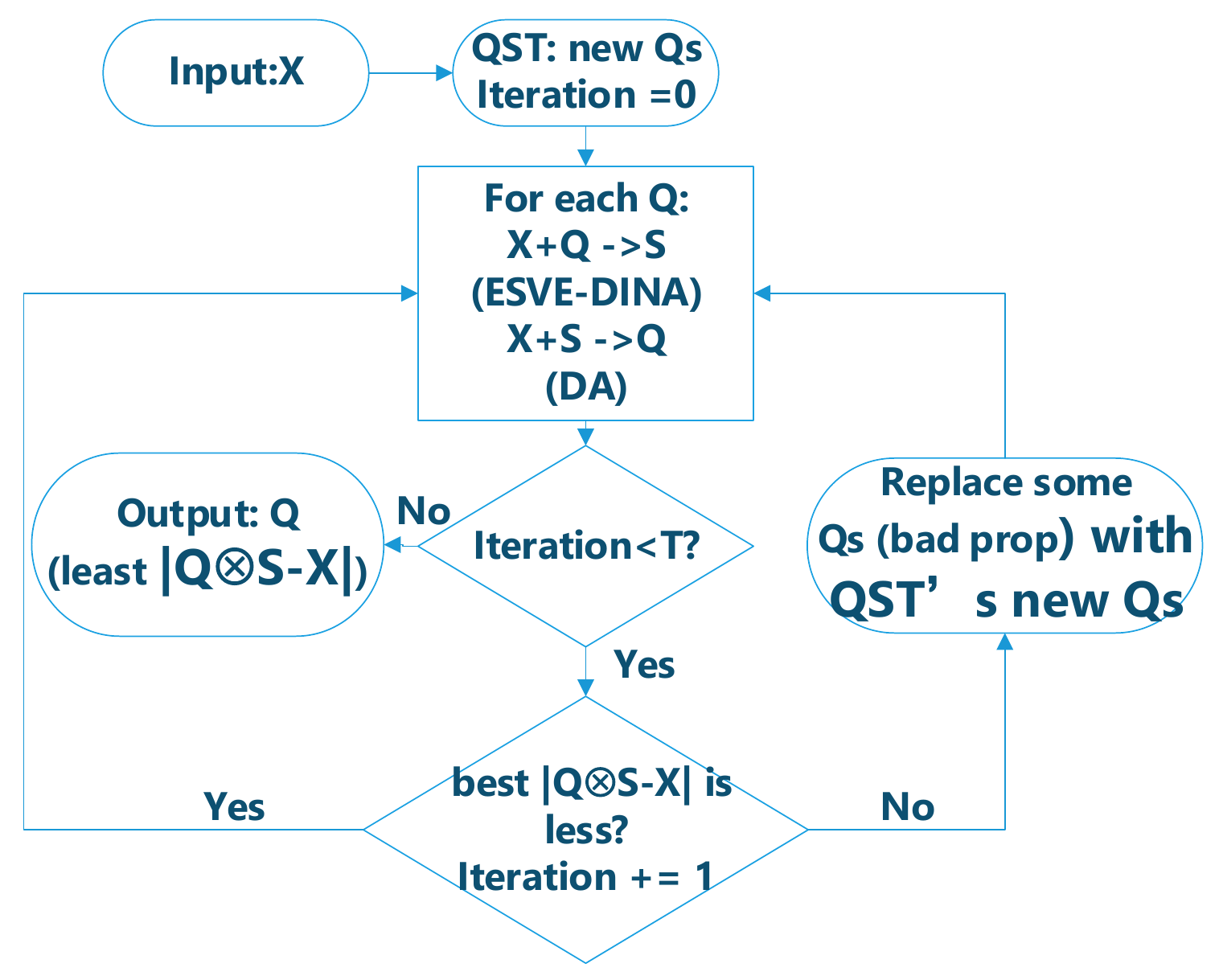}
		\caption{Framework of HBCA.}
	\end{figure}
	In summary, the HBCA framework is shown in Fig. 3, here, the $\bigotimes$ means the prediction method of optional specific models. 
	
	
	
	
	
	%
	\section{Experiments}
	In this section, first, we compare the performance of ESVE-DINA and HBCA against the baseline approaches (mainly the DINA model) on the PEP task. Next, we utilize some consistency experiments
	 to show that the student-independent s, g assumption of DINA is  inappropriate. Finally, we conduct experiments to investigate the hyperparameter  sensitivity of HBCA.
	
	
	\subsection{PEP Tasks}
	\subsubsection{Datasets}
	In our experiments, we adopted  \textit{Fraction}  \cite{tatsuoka1984analysis,wu2015cognitive,vie2019knowledge} and   \textit{ASSISTments2015} \cite{vie2019knowledge} (ASSIST for short) datasets in our experiments. 
	
	The Fraction dataset consists of scores of  fraction problems and
	 a  $Q$-matrix labeled by experts. The ASSIST dataset is collected by an online education system \cite{razzaq2005assistment}, and its $Q$-matrix is a sparse identity matrix. 
	We utilized the full Faction dataset and a part of the ASSIST dataset, because the original ASSIST dataset is sparse and contains duplicate records. We used three steps to select a subset, namely, filtering duplication, selecting questions with a record frequency of more than 20\% and selecting students with a response frequency of more than 50\%. Table 2 summarizes the data statistics of the selected datasets. Following \cite{wu2015cognitive, wang2019neural}, we then utilize 80\% of the  dataset chosen randomly for training and the remaining 20\% for testing. For additional comparison to  \cite{wu2015cognitive}, we also selected 20\%, 40\%, 60\% and 80\% of the dataset for testing on the Fraction dataset.
	
	
	\subsubsection{Experimental Setup}
	

	Among these experiments, ESVE-DINA has no hyperparameters. In HBCA, for some settings, we used the same values for the two datasets. We initialized 100 $Q$-matrices for 100 iterations and set the update number of the $Q$-matrix to 40. Its dual algorithm  used 100 random student vectors to estimate the $Q$-matrix.
	\begin{table}[htbp]
		\begin{center}
			\caption{\label{tab:test}Datasets used for the experiments.} 
			\begin{tabular}{lclcl} 
				\toprule 
				Dataset&Fraction& ASSIST\\ 
				\midrule 
				Students &  536 & 439  \\
				
				Exercises &  20 & 35  \\
				Skills &  8 & 35  \\
				Response logs &  10720 & 8389  \\
				Type of $Q$  & specific &  nonspecific \\
				\bottomrule 
			\end{tabular} 
		\end{center}
	\end{table}
	There are two differences between the two datasets; that is, the threshold $\eta$ in the Fraction dataset was 0.85, while that in ASSIST was 0.9, and their settings are discussed in \S 5.1. The HBCA searched the question vector dimension $dim_{qv}$ from 5 to 9 in the Fraction dataset, while ASSIST searched $dim_{qv}$ from 6 to 10 for more questions.
	All  of these methods are implemented on a Core i5 2.3GHz machine with a CPU. We built the following models for comparison:
	\begin{itemize}  
		\item \textit{ESVE-DINA-SI} 
		This model  utilizes the student vectors of ESVE-DINA and the
		 student-independent (\textbf{SI}) assumption of DINA. 
		\item \textit{ESVE-DINA-SD} 
		This model utilizes the student vectors of ESVE-DINA and our student-dependent (\textbf{SD}) s, g assumption that every $s_j$ and $g_j$ is  related to the student vectors.
		\item \textit{HBCA} 
		This  model uses the $Q$-matrix  labeled by HBCA instead of the original manual $Q$-matrix. HBCA will previously choose 20\% of the training set as validation to select a best $dim_{qv}$  with the corresponding selecting goal; i.e., x + HBCA means that the selecting goal is  MAE of x. 
		Since our API of the DINA code has a conflict with HBCA, DINA + HBCA uses Q of HBCA with the selecting goal of ESVE-DINA-SI instead, because they use the same SI s, g assumption.
		\item \textit{QST} 
		This model uses $Q$-matrix labeled by QST of the first iteration in  HBCA.
	\end{itemize}
	\subsubsection{Baselines}
	To demonstrate the effectiveness of ESVE-DINA and HBCA, we compare them with some baseline methods, their details are shown as follows:
	\begin{itemize}  
		\item \textit{DINA} \cite{junker2001cognitive,de2011generalized}: A model assumes that the probabilistic parameters of  questions are student-independent. Here, we implement DINA with the classic EM algorithm. DINA is a typical baseline \cite{wang2019neural,wu2015cognitive} for the PEP task.
		\item \textit{DINA} (Wu.) The results come  from a paper \cite{wu2015cognitive}.
		\item \textit{FuzzyCDF} \cite{wu2015cognitive}: A model that expands  each dimension of DINA  into an IRT model
		 \cite{rasch1961general}, it has more explainable but  complicated cognitive parameters than the DINA model. This is an enhanced comparison for our methods, and the numerical results here come from the original author.
		
	\end{itemize}
	
	
	\begin{table}[htbp]  \footnotesize
		\centering
		\caption{MAE comparison on the Fraction dataset.}  
		\label{tab:methodcompare}
		\setlength{\tabcolsep}{2mm}{
			\begin{tabular}{c|c|c|c|c}
				\hline
				\multirow{2}*{Method} & \multicolumn{4}{c}{Test ratio}  \\ 
				\cline{2-5}
				& 80\% & 60\% & 40\%   & 20\% \\
				\hline 
				DINA  &  0.5028 & 0.4457 & 0.3782& 0.3101  \\
				FuzzyCDF \cite{wu2015cognitive} &  0.3259 & 0.2867 & 0.2763 &\bf 0.2348\\
				ESVE-DINA-SD  &\bf  0.3137 &\bf 0.2672 &\bf 0.2497 & 0.2443 \\
				\hline
		\end{tabular}}
	\end{table}


	\subsubsection{Evaluation Metrics}
	
	To demonstrate the effectiveness of our models, we conduct experiments with five-time random validation on the PEP task, i.e., predicting response logs. We use the evaluation metrics from both classification aspect \cite{wu2015cognitive} and
	regression aspect \cite{wang2019neural}, including MAE (mean absolute error), RMSE (root mean square error) and AUC (area under the curve).  
	\subsubsection{Experimental Results}
	\begin{table*}[tp]\footnotesize
		\centering
		\caption{Experimental results on the predicting examinee performance task of the Fraction dataset.}  
		
		\scalebox{1}{  
			\setlength{\tabcolsep}{4mm}{
				\begin{tabular}{c|ccc}
					\hline
					Methods& MAE & RMSE & AUC  \\
					\hline 
					DINA (Wu.) \cite{wu2015cognitive} 
					& 0.3153 & 0.4056 & -\\
					DINA & 0.3101 & 0.3997 & 0.8577\\		
					%
					\hline
					ESVE-DINA-SI  & 0.2611 & 0.4595 & 0.7633  \\
					ESVE-DINA-SD  &\bf 0.2443 & 0.3865 & \bf 0.8704  \\
					DINA + HBCA & 0.3097 & 0.3903 & 0.8506\\
					
					ESVE-DINA-SI + HBCA  &   0.2556 &   0.4242 & 0.8032  \\
					ESVE-DINA-SD + HBCA  &  0.2561 &  \bf 0.3850 &  0.8649 \\
					\hline
		\end{tabular}}}
	\end{table*}
	
	\begin{table*}[tp]\footnotesize
		\centering
		\caption{Experimental results on the predicting examinee performance task of the ASSIST dataset.}  
		
		\scalebox{1}{  
			\setlength{\tabcolsep}{4mm}{
				\begin{tabular}{c|ccc}
					\hline
					Methods & MAE & RMSE & AUC   \\
					\hline 
					DINA    & 0.5101 & 0.5579 & 0.6505\\
					
					%
					\hline
					
					DINA + HBCA & 0.4961 & 0.5385 & 0.6428\\
					
					ESVE-DINA-SI + HBCA  &   0.4366 &   0.5813 & 0.5859  \\
					ESVE-DINA-SD + HBCA  &  \bf 0.3770 &  \bf 0.4707 & \bf 0.6691 \\
					\hline
		\end{tabular}}}
	\end{table*}
	Table 2 presents results with different test ratios on the Fraction dataset, and Tables 3 and 4  respectively show the results of our methods and other baseline approaches with 20\% test ratio  on the  Fraction and ASSIST datasets. 
	From Tables 3 and 4, we can observe that ESVE-DINA-SD outperforms DINA and ESVE-DINA-SI  with experts' $Q$ on Fraction dataset and Q of HBCA on both datasets, indicating the effectiveness of ESVE-DINA-SD. For HBCA,
	 different models mainly have differences in the selecting goals  but have the same initialization and calibration methods, so the results with Q of HBCA are approximately comparable. 
	 Second, comparing the same models with different $Q$-matrices, we can observe that the $Q$-matrix automatically labeled by HBCA can achieve comparable performance to that of the manual $Q$-matrices. Specifically, Q of HBCA shows similar results with DINA on both datasets, better results with ESVE-DINA-SI and slightly worse results with ESVE-DINA-SD on the Fraction dataset.
	Third, according to  Table 2, our ESVE-DINA-SD model shows better MAE than FuzzyCDF when the test ratio is more than 20\%, indicating that our model has stronger prediction ability with less student information.
	
	One detail is that ESVE-DINA can not solve ASSIST dataset, because its trivial unit  $Q$-matrix cannot be used for the conflict detection of ESVE-DINA.
	\subsection{Consistency Test of the s, g assumption of DINA}
	
	To examine the assumption of DINA that the slip and guess rates of each question are student-independent (SI),
	we compute  $s$ on different levels  with the golden student results of the test set and solved student vectors of training set as their references:
	\begin{align}
	s_{j,k}^{ref} = \frac{ \sum_{i=0}^{S} ((X_{ij}=0)  \& (\xi_{ij}=1) \& (sum(\boldsymbol \alpha_i) =k))}{\sum_{i=0}^{S} (  (\xi_{ij}=1) \& (sum(\boldsymbol  \alpha_i) =k))}
	\end{align}
	\begin{figure}
		\centering
		\subfigure[$s^{ref}$ heatmap of DINA.]{
			\scalebox{1}{  
				\includegraphics[width=1.54in]{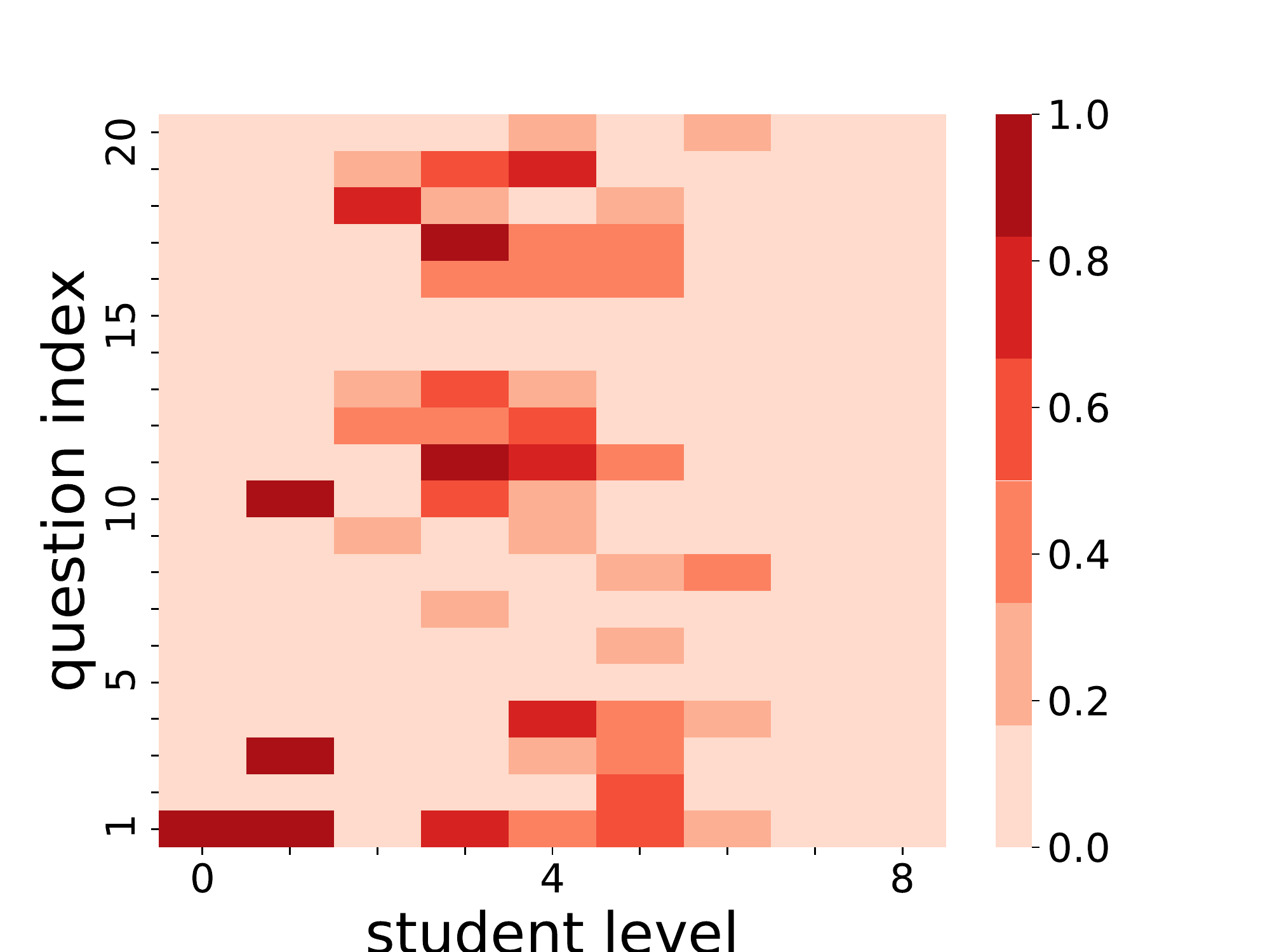}
			}
		}
		\hspace{-4mm} 
		\subfigure[$s_\delta, g_\delta$ comparison.]{
			\scalebox{1}{ 
				\includegraphics[width=1.54in]{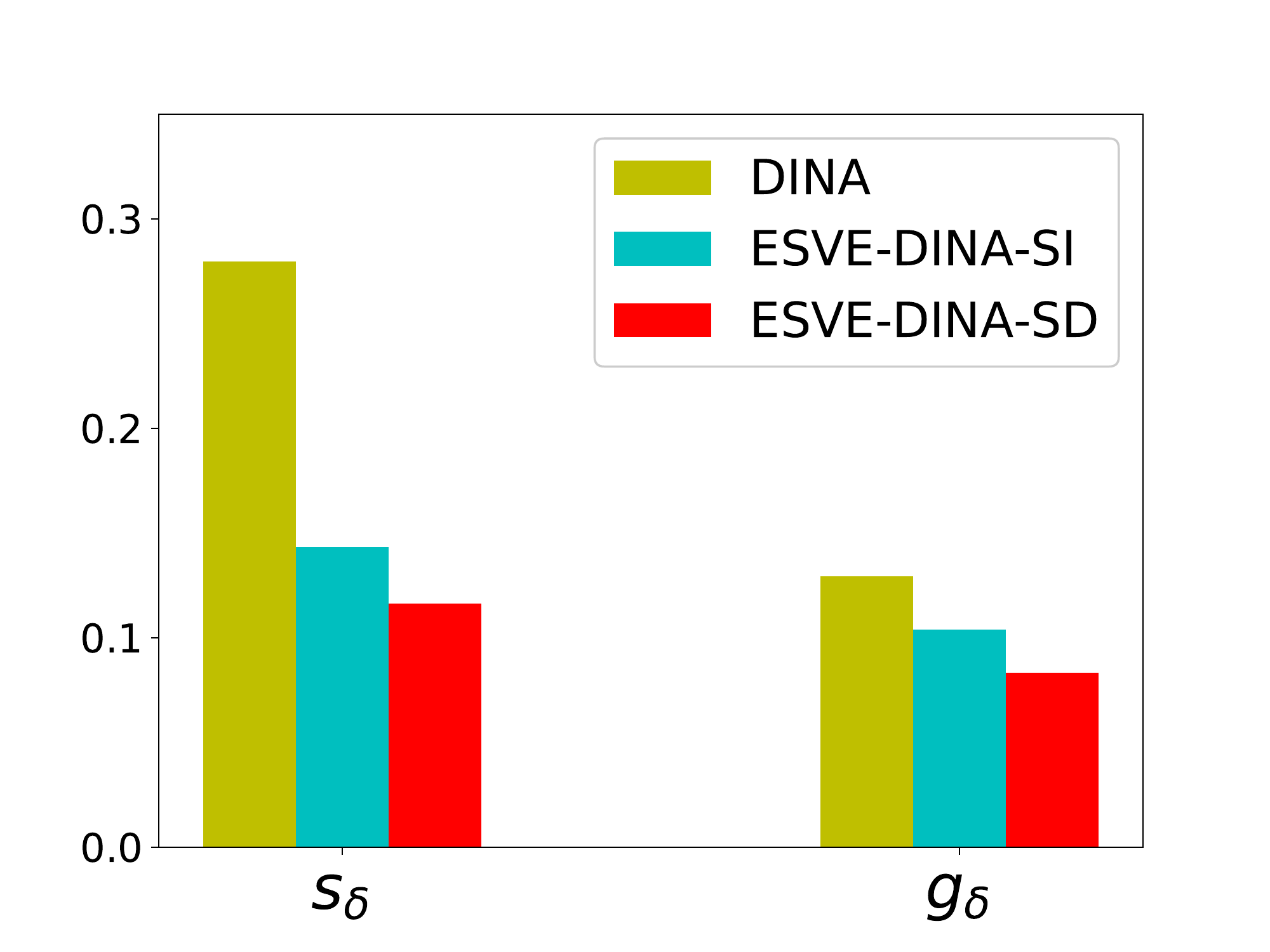}}
		}
		\caption{ Results of consistency experiments for examining the SI assumption of DINA.}
		\vspace{-1mm}
	\end{figure}
	Here,    sum($\boldsymbol  \alpha_i$) represents the level of student $i$. References of $g$ are similar, the level becomes deficiency, and the values of $X_{ij}, \xi_{ij}$ are reversed.
	Then, as the golden student results X of test set are unseen, the values of $s^{ref}, g^{ref}$ can be  references values of $s, g$. The distribution of $s^{ref}, g^{ref}$ can be validated to determine whether $s, g$  can be SI.
	Fig. 4 (a) shows the distribution of $s^{ref}$ of DINA  with experts' Q on the Fraction dataset, the results with Q of HBCA on both datasets or the results of $g^{ref}$ are similar. 
	 We can observe that the values of most rows have large variances, implying that the slip rates can be related to student levels.  
	Thus, the SI s, g assumption of DINA can be inappropriate.
	
	
	Next, we compare the distortions between estimated $s, g$ of the training set and their references  to demonstrate  the rationality comparison between the SI s,g  and SD s, g assumption. The distortion of $s$ is defined as follows (distortion of $g$ is similar):
	\begin{align}
	s_{\delta}=  \frac{1}{MN}\sum_{j=0}^M\sum_{k=0}^N |s_{j,k} - s_{j,k}^{ref}|.
	\end{align}
	
	Here, every $s_{j, k}$ is estimated on the training set, and $s_{j,k}^{ref}$ shown in Eq. (6.18) are references values of  the test set. Then, their consistency can  partly show the rationality of  s, g assumption. Moreover, for  ESVE-DINA-SI and DINA,  their SI $s_{j,k}$ values of the training set are constant on dimension $k$. Fig. 4 (b) shows the $s_\delta, g_\delta$ comparison with experts' Q on the Fraction dataset, the results with Q of HBCA on both datasets are similar.
	We can observe that our SD s, g of ESVE-DINA-SD has  smallest $s_\delta, g_\delta$. 
	Thus,  our SD s, g assumption is more reasonable than SI s, g assumption.
	\subsection{Hyperparameter Sensitivity of HBCA}
	%
	%
	
	Here, we show the hyperparameter sensitivity of our  unsupervised labeling method HBCA.  We test two primary factors, namely, the initialization (QST) efficiency and the  question vector dimension ($dim_{qv}$). 
	Fig. 6 shows the  MAE comparison of HBCA and QST on the validation and test sets  of the Fraction dataset, and similar results are obtained for the  ASSIST dataset. From Fig. 5, we can observe that the initialization of HBCA (QST) is  not bad, and it ($dim_{qv}=9$) is only
	slightly  worse than the results with experts' $Q$ (0.2443)  on the test set. Second, 
	 we  observe that labeling of HBCA is not sensitive to $dim_{qv}$, and it has relatively smooth  results on the  test set. Third, we also observe that the calibration results of HBCA (gap of HBCA and QST) are different  on the validation and test sets. This result may be due to the difference between the test pattern and labelling progress, because there is an entire training set to estimate parameters when testing a model, but the labeling progress selects the $Q$-matrix with the best performance of the validation set, which is only a subset of the original training set. In summary, we can observe that the labeling progress of HBCA is efficient due to its good initialization and dimensional robustness.
	
	

	\begin{figure}[htbp]
		\centering
		\includegraphics[scale=0.3]{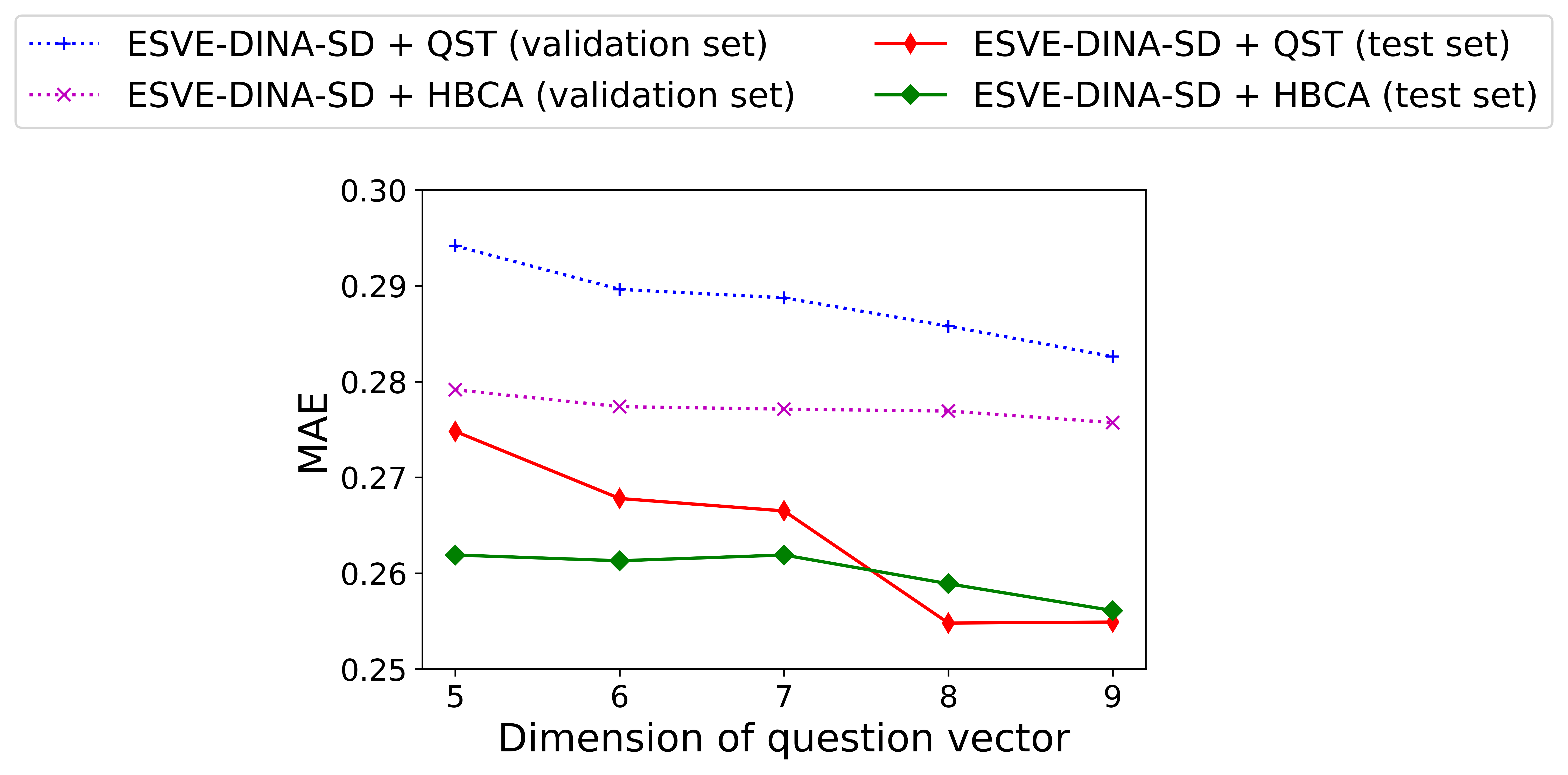}
		\caption{Parameter sensitivity of HBCA.}
	\end{figure}
	\section{Conclusion and Future Work}
	In this paper, we propose an ESVE algorithm to estimate  student vectors  of DINA  without any probabilistic assumption outside its cognitive assumption, and we design student-dependent slip and guess rates with estimated student vectors, which  avoids the strong student-independent assumption of DINA. 
	Moreover, we design an unsupervised  method  HBCA to label the $Q$-matrix automatically based on the  question relation and the fault tolerance of ESVE-DINA. Finally, experiments on two real-world datasets show that ESVE-DINA  outperforms the original DINA model on  accuracy, and the $Q$-matrix automatically labeled by HBCA can achieve  performance comparable to that of  the manual $Q$-matrix when using the same model structure. 
	
	In future work, there are still some further studies. First,  a better probabilistic model may improve ESVE-DINA, because our student-dependent assumption only uses the count for the parameter estimation. Second, the initialization of the $Q$-matrix in HBCA is unrelated to DINA,
 and it may be further investigated for  initializing the $Q$-matrix of all CDMs.
	Third, estimating the $Q$-matrix from the estimated student information may be a novel approach for other CDMs.
	\bibliographystyle{siam}
	\bibliography{dina_bib}
	%
	%
	%
	
\end{document}